\documentclass[prd,aps,showpacs,preprintnumbers,amssymb]{revtex4}
\usepackage{axodraw}
\usepackage{color}
\usepackage{epsf}
\usepackage{graphicx}
\def\e3p{$\eta \rightarrow 3 \pi$}

\begin{document}
\title{%
\hfill{\normalsize\vbox{%
\hbox{}
 }}\\
{Note about lepton masses and mixings in two Ansatze}}
\author{Renata Jora
$^{\it \bf a}$~\footnote[1]{Email:
 rjora@theory.nipne.ro}}

\author{Joseph Schechter
 $^{\it \bf b}$~\footnote[2]{Email:
 schechte@phy.syr.edu}}

\author{M. Naeem Shahid
$^{\it \bf c}$~\footnote[3]{Email:
   mnshahid@phy.syr.edu }}

\affiliation{$^{\bf \it a}$ National Institute of Physics and Nuclear Engineering PO Box MG-6, Bucharest-Magurele, Romania}
\affiliation{$^ {\bf \it b}$ Department of Physics,
 Syracuse University, Syracuse, NY 13244-1130, USA}
\affiliation{$^ {\bf \it c}$ Department of Physics, School of Natural Sciences (SNS),
National University of Sciences and Technology (NUST), H-12, Islamabad, Pakistan}

\date{\today}

\begin{abstract}
We consider two Ansatze for the neutrino masses and mixings in which the permutation symmetry is implemented in various orders.
We discuss the possible see-saw mechanisms and the charged lepton masses for the two cases in the presence of a Higgs triplet and  three Higgs doublets.
\end{abstract}
\maketitle

\section{Introduction}

The problem of neutrino masses and mixings is at the forefront of experimental research \cite{K}-\cite{MINOS} in particle physics and has inspired many theoretical models \cite{Wolfenstein}-\cite{Jora4} involving continuous or discrete symmetries
in the flavor sector of the standard model.

If $U$ is the neutrino mixing matrix and $W$ is the charged lepton mixing matrix then one of the main experimental results involves the product $K=W^{\dagger}U$, the lepton mixing matrix. Many studies suggest that at least in first approximation $W=1$ such that $K=U$.

From neutrino oscillations experiments we learn that \cite{Nakamura}:
\begin{eqnarray}
&&A\equiv m_2^2-m_1^2=(7.50\pm0.20)\times 10^{-5} {\rm eV}^2
\nonumber\\
&&B\equiv|m_3^2-m_2^2|=(2.32^{+0.12}_{-0.08})\times 10^{-3} {\rm eV}^2
\label{res4443}
\end{eqnarray}

These results suggest two possibilities depending on the magnitude of $m_3$:
\begin{eqnarray}
&& Type\, I \,\,\,{\rm normal\,hierarchy}\,\,\, m_3>m_2>m_1
\nonumber\\
&& Type\, II\,\,\,{\rm inverted\, hierarchy}\,\,\,m_2>m_1>m_3.
\label{res332221}
\end{eqnarray}

In this paper, we will discuss the see-saw mechanisms for neutrino masses and mixings in two contexts, in both assuming an $S_3$ symmetry and  its perturbations in the lepton sector of the standard model.

\section{An Ansatz for Dirac neutrino masses and mixings}

We consider a Dirac mass matrix for the neutrinos which is diagonalized by the unitary matrix $U$:
\begin{eqnarray}
U=
\left[
\begin{array}{ccc}
u_{11}&u_{12}&u_{13}\\
u_{21}&u_{22}&u_{23}\\
u_{31}&u_{32}&u_{33}
\end{array}
\right].
\label{matr55546}
\end{eqnarray}

In the most general case the above matrix diagonalizes the hermitian form  $M=M_1M_1^{\dagger}$:
\begin{eqnarray}
U^{\dagger}MU=M_d^2,
\label{res554546}
\end{eqnarray}

where $M_1$ is the standard Dirac neutrino mass matrix:

The following discussion applies both to the case of type I (where $m_3$ is the largest mass) and II ( where $m_3$ is the smallest mass) neutrinos:

We can write:
\begin{eqnarray}
&&M=UM_d^2U^{\dagger}
\nonumber\\
&&M=
\left[
\begin{array}{ccc}
u_{11}&u_{12}&u_{13}\\
u_{21}&u_{22}&u_{23}\\
u_{31}&u_{32}&u_{33}
\end{array}
\right]
\left[
\begin{array}{ccc}
m_{d1}^2&0&0\\
0&m_{d2}^2&0\\
0&0&m_{d3}^2
\end{array}
\right]
\left[
\begin{array}{ccc}
u_{11}^*&u_{21}^*&u_{31}^*\\
u_{12}^*&u_{22}^*&u_{32}^*\\
u_{13}^*&u_{23}^*&u_{33}^*
\end{array}
\right]
\label{res44398}
\end{eqnarray}

This would lead to the following decomposition of the most general general hermitian form of the mass matrix:
\begin{eqnarray}
M=m_{d1}^2
\left[
\begin{array}{ccc}
u_{11}u_{11}^*&u_{11}u_{21}^*&u_{11}u_{31}^*\\
u_{21}u_{11}^*&u_{21}u_{21}^*&u_{21}u_{31}^*\\
u_{31}u_{11}^*&u_{31}u_{21}^*&u_{31}u_{31}^*
\end{array}
\right]
+m_{d2}^2
\left[
\begin{array}{ccc}
u_{12}u_{12}^*&u_{12}u_{22}^*&u_{12}u_{32}^*\\
u_{22}u_{12}^*&u_{22}u_{22}^*&u_{22}u_{32}^*\\
u_{32}u_{12}^*&u_{32}u_{22}^*&u_{32}u_{32}^*
\end{array}
\right]
+m_{d3}^2
\left[
\begin{array}{ccc}
u_{13}u_{13}^*&u_{13}u_{23}^*&u_{13}u_{33}^*\\
u_{23}u_{13}^*&u_{23}u_{23}^*&u_{23}u_{33}^*\\
u_{33}u_{13}^*&u_{33}u_{23}^*&u_{33}u_{33}^*
\end{array}
\right].
\label{eq33245}
\end{eqnarray}

However, it would be useful to know the initial form of the mass matrix $M_1$ from which one can extract the hermitian form of $M=M_1M^{\dagger}_1$.
For that we assume that,
\begin{eqnarray}
M_1=m_{d1}
\left[
\begin{array}{ccc}
u_{11}&0&0\\
u_{21}&0&0\\
u_{31}&0&0
\end{array}
\right]
+m_{d2}
\left[
\begin{array}{ccc}
0&u_{12}&0\\
0&u_{22}&0\\
0&u_{32}&0
\end{array}
\right]
+m_{d3}\left[
\begin{array}{ccc}
0&0&u_{13}\\
0&0&u_{23}\\
0&0&u_{33}
\end{array}
\right].
\label{rt555}
\end{eqnarray}
 then,

\begin{eqnarray}
M=M_1M^{\dagger}_1.
\label{res4443}
\end{eqnarray}

\section{Short review of an $S_3$ symmetric model}

In  \cite{Jora1} we proposed a model for the neutrino masses and mixings which involved the implementation of the $S_3$ symmetry into the lepton sector of the standard model.
This model was further studied in \cite{Jora2}, \cite{Jora3}, \cite{Jora4} where perturbation symmetric under $S_2$ subgroups of $S_3$ were applied and also a possible CP violation phase was considered.
We start with an $S_3$ zeroth order symmetric mass matrix for the Majorana neutrinos:
\begin{eqnarray}
M_{\nu}^0=
\alpha
\left[
\begin{array}{ccc}
1&0&0\\
0&1&0\\
0&0&1
\end{array}
\right]
+
\beta
\left[
\begin{array}{ccc}
1&1&1\\
1&1&1\\
1&1&1
\end{array}
\right]=\alpha \textbf{1} +\beta d
\label{fsirst55466}
\end{eqnarray}

This matrix is diagonalized to:
\begin{eqnarray}
R^TM_{\nu}^0R=
\left[
\begin{array}{ccc}
\alpha&0&0\\
0&\alpha+3\beta&0\\
0&0&\alpha
\end{array}
\right],
\label{diag665}
\end{eqnarray}

where $R$ is the so called tribimaximal form,
\begin{eqnarray}
R=
\left[
\begin{array}{ccc}
-2/\sqrt{6}&1/\sqrt{3}&0\\
1/\sqrt{6}&1/\sqrt{3}&1/\sqrt{2}\\
1/\sqrt{6}&1/\sqrt{3}&-1/\sqrt{2}
\end{array}
\right].
\label{tr768}
\end{eqnarray}

Since in zeroth order the neutrino masses can be approximated as degenerate we required for $\beta$ to be real and also,
\begin{eqnarray}
&&\alpha\equiv-i|\alpha|e^{-i\Psi/2}
\nonumber\\
&&|\alpha|=\frac{3\beta}{2\sin(\Psi/2)}.
\label{rel88768}
\end{eqnarray}

In order to obtain a better agreement with the experimental data, we then added the first order perturbation:
\begin{eqnarray}
\Delta=
\left[
\begin{array}{ccc}
0&0&0\\
0&t&u\\
0&u&t
\end{array}
\right],
\label{fir6657}
\end{eqnarray}

and a second order perturbation of the type:
\begin{eqnarray}
\Delta'=
\left[
\begin{array}{ccc}
t'&u'&0\\
u'&t'&0\\
0&0&0
\end{array}
\right].
\label{pert5546}
\end{eqnarray}

We made the underlying assumption that the charged lepton mixing matrix is the unit matrix. The $S_3$ symmetry is implemented by  three Higgs doublets and the left handed charged leptons, both of them  in the three dimensional representation of the symmetric group. The right handed charged leptons are considered singlets.  Then the effective Lagrangian has the form:
\begin{eqnarray}
{\cal L}=-\sqrt{2} \sum_{c,b}\bar{e}_{Rc}\sum_a\Phi_a^{\dagger}G_{ab}^{(c)}L_b+ h.c.,
\label{eff453}
\end{eqnarray}

Here $G^{(c)}$ is invariant under the $S_3$ transformation,
\begin{eqnarray}
G^{(c)}=\gamma^{(c)}\textbf{1}+\delta^{(c)}d
\label{ret77}
\end{eqnarray}

and $\gamma^{(c)}$ and $\delta^{(c)}$ are complex numbers.

Then the mass matrix is given by:
\begin{eqnarray}
M_e=
\left[
\begin{array}{ccc}
v_1\gamma^{(1)}+\lambda\delta^{(1)}&v_2\gamma^{(1)}+\lambda\delta^{(1)}&v_3\gamma^{(1)}+\lambda\delta^{(1)}\\
v_1\gamma^{(2)}+\lambda\delta^{(2)}&v_2\gamma^{(2)}+\lambda\delta^{(2)}&v_3\gamma^{(2)}+\lambda\delta^{(2)}\\
v_1\gamma^{(3)}+\lambda\delta^{(3)}&v_2\gamma^{(3)}+\lambda\delta^{(3)}&v_3\gamma^{(3)}+\lambda\delta^{(3)}
\end{array}
\right]
\label{res44343}
\end{eqnarray}
and one can obtain the correct spectrum for a suitable choice of the parameters $\gamma^{(c)}$ and $\delta^{(c)}$. Here $\lambda=v_1+v_2+v_3$.

\section{A see-saw mechanism for an $S_3$ symmetric model}

As mentioned in section III the model involves a Higgs triplet which couples to the neutrinos and three Higgs doublets which couple with the charged leptons. Note that the presence of a Higgs triplet and of multiple Higgs doublets is natural in the context of $SO(10)$ GUT theories where the Higgs triplet comes from the (10) dimensional representation
and the Higgs doublets can come from the $(10)$, $(120)$ or  $(126)$ representations.

We assume that the smallness of the neutrino masses is due to the heavy Higgs triplet with mass $M$ which couples with the Higgs doublet $k$ with the strength $\mu_k$.
Then by integrating out the Higgs triplet one obtains for the Majorana neutrino mass:
\begin{eqnarray}
M_{\nu}=-\frac{f_{ij}}{M^2}\sum_k(\mu_k v_k^2),
\label{neutr778}
\end{eqnarray}
where $f_{ij}$ is the coupling of $\nu_i$, $\nu_j$ with the Higgs triplet and $v_k$ is the vacuum expectation value of the $k^{th}$ ($k=1,2,3$) Higgs doublet.
If we want to assume the invariance under $S_3$ of the Lagrangian we need universal coupling $\mu_K=\mu$ since the three Higgs doublet are in the fundamental representation of $S_3$.
Then Eq. (\ref{neutr778}) becomes:
\begin{eqnarray}
M_{\nu}=-\frac{f_{ij}}{M^2}\mu(\sum_k v_k^2)
\label{rexs44}
\end{eqnarray}

This mass matrix is invariant under $S_3$ at zeroth order and its perturbations are invariant under the subgroups, $S_2$ of $S_3$ (see section III).

The charged lepton masses are obtained as in section III by couplings with the three Higgs doublets.

\section{ See-saw mechanism with right handed neutrinos}

We consider a model with a Higgs triplet, three Higgs doublets and we add three right handed heavy neutrino $N_{kR}$ ($k=1,2,3$) with a Majorana mass.
We further consider that the couplings of the three neutrinos with the Higgs triplet are universal and the corresponding mass matrix is diagonal.

Assume we arrange both right handed neutrino and the left handed neutrinos in a three dimensional representation of the group $S_3$. We also set the three Higgs doublets in  a $3\times3 $ diagonal matrix,
\begin{eqnarray}
\Phi=
\left[
\begin{array}{ccc}
\Phi_1&0&0\\
0&\Phi_2&0\\
0&0&\Phi_3
\end{array}
\right],
\label{arr54677}
\end{eqnarray}

that transforms under $S_3$ as,
\begin{eqnarray}
\Phi'=S\Phi S^{\dagger}.
\label{rez5647}
\end{eqnarray}

Then a coupling term among the left handed, right handed neutrinos and the Higgs doublets that is invariant under the $S_3$ symmetry has the form:
\begin{eqnarray}
{\cal L}_n=y\bar{N}_{R}\Phi\Psi_L+h.c.
\label{r546}
\end{eqnarray}

where y is an arbitrary dimensionless constant.

An $S_3$ symmetric Majorana mass term for the right handed neutrinos is realized through the Lagrangian:
\begin{eqnarray}
{\cal L}_R=-\frac{1}{2}\bar{N}_{i}M_{ij}N_j,
\label{right89}
\end{eqnarray}

where the mass matrix M is of the type:
\begin{eqnarray}
M=a{\bf 1}+b d=U^*M_dU^{\dagger}
\label{q234},
\end{eqnarray}

and it is diagonalized by an arbitrary tribimaximal matrix U with two degenerate eigenvalues and one different.
In the $N_k'$ mass eigenstate basis the term ${\cal L}_n$ will become:
\begin{eqnarray}
{\cal L}_n=y\bar{N}'_RU^{\dagger}\Phi\Psi_L
\label{lagr5674}
\end{eqnarray}

Note that this Dirac mass term realizes the Ansatz in section II.

Upon integrating out the heavy right handed neutrino fields one obtain a Majorana mass for the light neutrinos of the type:
\begin{eqnarray}
m=
y^2\left[
\begin{array}{ccc}
v_1&0&0\\
0&v_2&0\\
0&0&v_3
\end{array}
\right]
 U^{*}M^{-1}_{d}U^{\dagger}
\left[
\begin{array}{ccc}
v_1&0&0\\
0&v_2&0\\
0&0&v_3
\end{array}
\right],
\label{mass67589}
\end{eqnarray}

which can be further simplified  by stating:
\begin{eqnarray}
 U^{*}M^{-1}_{d}U^{\dagger}=M^{-1}.
\label{expr77890}
\end{eqnarray}

We shall try implement the permutation group $S_3$ and perturbations of it into the Ansatz in Eq. (\ref{mass67589}).

First we decompose the matrix of vacuum expectation values as follows:
\begin{eqnarray}
\left[
\begin{array}{ccc}
v_1&0&0\\
0&v_2&0\\
0&0&v_3
\end{array}
\right]
=
\left[
\begin{array}{ccc}
v_1-v_2+v_3&0&0\\
0&v_1-v_2+v_3&0\\
0&0&v_1-v_2+v_3
\end{array}
\right]
+
\left[
\begin{array}{ccc}
v_2-v_3&0&0\\
0&v_2-v_3&0\\
0&0&0
\end{array}
\right]
+
\left[
\begin{array}{ccc}
0&0&0\\
0&v_2-v_1&0\\
0&0&v_2-v_1
\end{array}
\right].
\label{deomp78}
\end{eqnarray}

We  denote:
\begin{eqnarray}
&&v_1-v_2+v_3=k
\nonumber\\
&&v_2-v_3=t
\nonumber\\
&&v_2-v_1=u
\label{rez5647}
\end{eqnarray}

We consider the Majorana contribution to the masses coming from the coupling with the Higgs triplet given by $m_0$ and proportional to the unit matrix in the flavor space and ask:
\begin{eqnarray}
m_0+y^2k^2M^{-1}\,\,{\rm\,of\,zeroth\,order}.
\label{rez4536}
\end{eqnarray}

Since the inverse of  matrix $M$ is $M^{-1}$ given by,
\begin{eqnarray}
M^{-1}=\frac{1}{a(a+3b)}[(a+3b){\bf 1}-b d]
\label{inv6758}
\end{eqnarray}

 we have in zeroth order  a neutrino mass term invariant under $S_3$.

In first order the contribution is given by,
\begin{eqnarray}
&&y^2M^{-1}
\left[
\begin{array}{ccc}
t&0&0\\
0&t&0\\
0&0&0
\end{array}
\right]
+y^2\left[
\begin{array}{ccc}
t&0&0\\
0&t&0\\
0&0&0
\end{array}
\right]M^{-1}+y^2
\left[
\begin{array}{ccc}
t&0&0\\
0&t&0\\
0&0&0
\end{array}
\right]M^{-1}
\left[
\begin{array}{ccc}
t&0&0\\
0&t&0\\
0&0&0
\end{array}
\right].
\label{rez5647}
\end{eqnarray}

and this term is invariant under the $S_{12}$ subgroup of $S_3$ as each of its factors is invariant.   We consider these terms as belonging to the first order perturbation.
Then similarly the second order term is invariant under an $S_{23}$ subgroup of $S_3$.

We arrange the left handed and right handed charged leptons in the three dimensional representation of the group $S_3$.
The relevant Lagrangian for the charged lepton masses is:
\begin{eqnarray}
{\cal L}=-\sqrt{2}\bar{e}_{Rk}M^{e1}_{ki}\Phi^{\dagger}_{ii}M^{e2}_{im}L_m
\label{ch66657}
\end{eqnarray}

and it is invariant  under the $S_3$ symmetry provided that:
\begin{eqnarray}
&&M^{e1}=\gamma^1{\bf 1} +\delta^{1}d
\nonumber\\
&&M^{e2}=\gamma^2{\bf 1}+\delta^2 d
\label{res4467435}
\end{eqnarray}

In first order we consider these matrices proportional to the unit matrix such that the hierarchy of the charged lepton masses is given by the hierarchy of the three Higgs vacuum expectation values. Then from the known hierarchy of the charged leptons and the electroweak vacuum,
\begin{eqnarray}
&&\frac{m_{e}}{m_{\mu}}=\frac{v_1}{v_2}
\nonumber\\
&&\frac{m_e}{m_{\tau}}=\frac{v_1}{v_3}
\nonumber\\
&&v_1^2+v_2^2+v_3^2=v^2=246^2 \,\,{\rm GeV^2}
\label{vc6578}
\end{eqnarray}

one obtains the vacuum expectation values of the three Higgs doublets:
\begin{eqnarray}
&&v_1=0.078 \,\,{\rm GeV}
\nonumber\\
&&v_2=14.62\,\, {\rm GeV}
\nonumber\\
&&v_3=245.8\,\,{\rm GeV}.
\label{rez6758}
\end{eqnarray}

For consistency we require that the masses of the heavy right handed neutrinos are degenerate and have the value $m$ in accordance with the picture depicted in section III.
Moreover we either set the parameter $y=1$ or absorb it in m.
Thus in zeroth order the neutrino masses are:
\begin{eqnarray}
m_{\nu}=m_0+\frac{k^2}{m}.
\label{zer647}
\end{eqnarray}

Constraints on cosmological structure formation \cite{Spergel}-\cite{Han} lead to a rough bound on the new neutrino masses:
\begin{eqnarray}
m_{\nu 1}+m_{\nu 2}+m_{\nu 3}\leq 0.3 \,\,{\rm eV}
\label{con768}
\end{eqnarray}

such that in the zeroth order we take,
\begin{eqnarray}
m_0+\frac{k^2}{m}\simeq 0.1 \,{\rm eV}
\label{lim87968}
\end{eqnarray}

which further yields: $m\geq 0.53\times 10^{15}$ GeV.

We denote:
\begin{eqnarray}
&&t_1=\frac{tk/m}{m_0+k^2/m}
\nonumber\\
&&t_2=\frac{t^2/m}{m_0+k^2/m}
\nonumber\\
&&u_1=\frac{uk/m}{m_0+k^2/m}
\nonumber\\
&&u_2=\frac{u^2/m}{m_0+k^2/m}.
\label{not8790}
\end{eqnarray}

\begin{figure}
\begin{center}
\epsfxsize = 10 cm
 \epsfbox{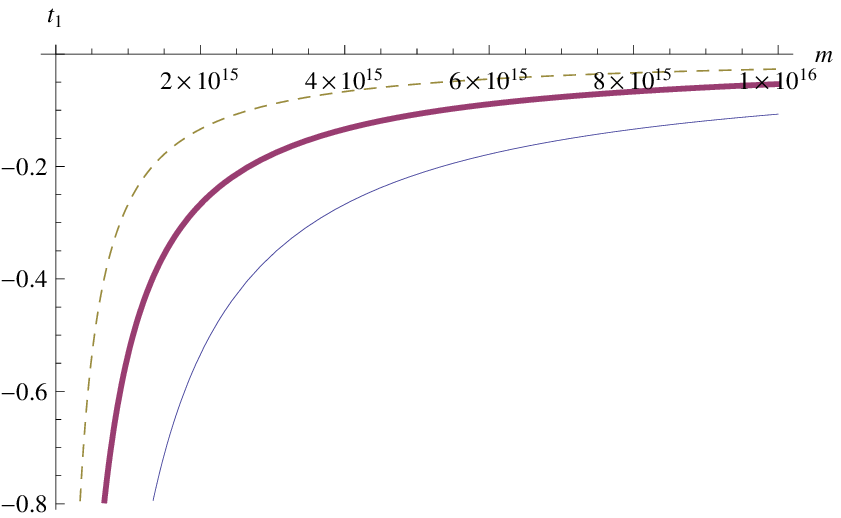}
\end{center}
\caption[]{%
Plot of the parameter $t_1$ as function of the mass of the heavy right handed neutrinos. The thick line corresponds to a zeroth order neutrino mass ($m_{\nu}$) of 0.1 eV, the thin line to  $m_{\nu}=0.05$ eV and the dashed line to $m_{\nu}=0.2$ eV.

}
\label{a1}
\end{figure}
\begin{figure}
\begin{center}
\epsfxsize = 10 cm
 \epsfbox{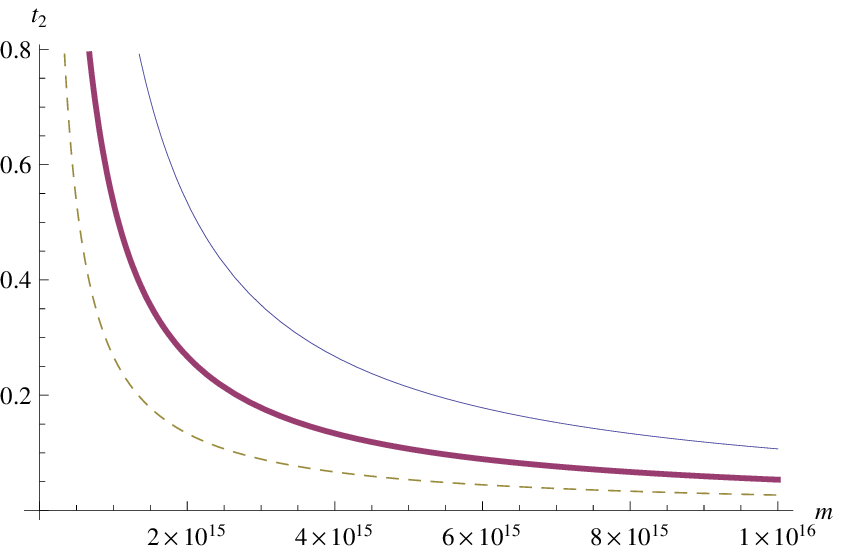}
\end{center}
\caption[]{%
Plot of the parameter $t_2$ as function of the mass of the heavy right handed neutrinos. The thick line corresponds to a zeroth order neutrino mass ($m_{\nu}$) of 0.1 eV, the thin line to  $m_{\nu}=0.05$ eV and the dashed line to $m_{\nu}=0.2$ eV.
}
\label{a2}
\end{figure}
\begin{figure}
\begin{center}
\epsfxsize = 10 cm
 \epsfbox{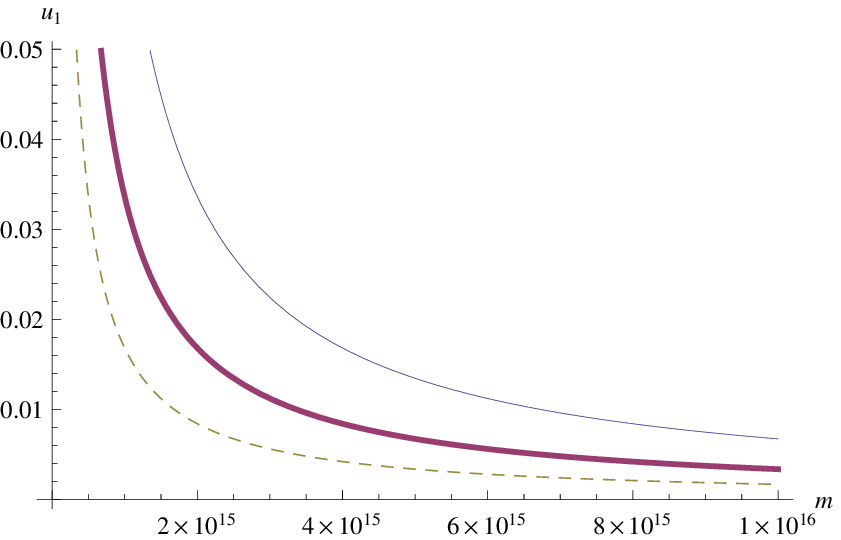}
\end{center}
\caption[]{%
Plot of the parameter $u_1$ as function of the mass of the heavy right handed neutrinos. The thick line corresponds to a zeroth order neutrino mass ($m_{\nu}$) of 0.1 eV, the thin line to  $m_{\nu}=0.05$ eV and the dashed line to $m_{\nu}=0.2$ eV.
}
\label{a3}
\end{figure}
\begin{figure}
\begin{center}
\epsfxsize = 10 cm
 \epsfbox{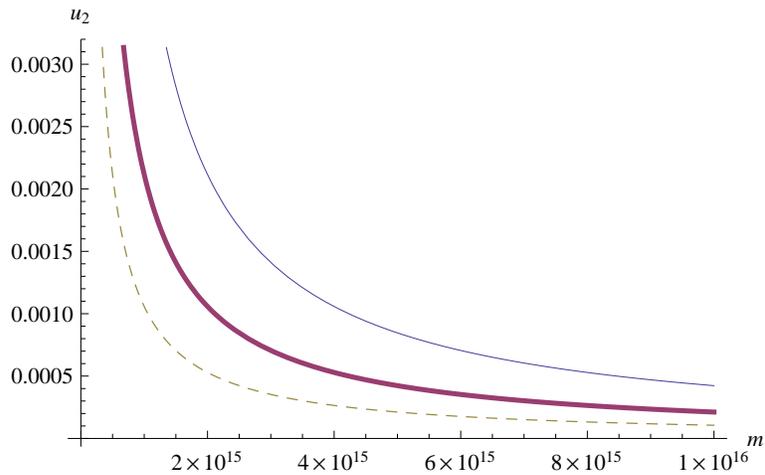}
\end{center}
\caption[]{%
Plot of the parameter $u_2$ as function of the mass of the heavy right handed neutrinos. The thick line corresponds to a zeroth order neutrino mass ($m_{\nu}$) of 0.1 eV, the thin line to  $m_{\nu}=0.05$ eV and the dashed line to $m_{\nu}=0.2$ eV.
}
\label{a4}
\end{figure}

In Figs. \ref{a1}, \ref{a3} we plot the parameters $t_1$ and $t_2$ in terms of the heavy mass of the right handed neutrinos for three possible zeroth order values of the mass of the light neutrinos. In Figs. \ref{a2}, \ref{a4} we plot the parameters $u_1$ and $u_2$ for the same cases.

A typical point for example is for $m=1.78\times 10^{15}$ GeV where $t_1=-0.3$, $t_2=0.3$ and corresponds to the first order perturbation whereas $u_1=0.02$ and corresponds to a second order perturbation. The term in the Lagrangian of order $u_2=0.0011$ is small and can be neglected.

\section{Conclusions}

The implementation of the symmetric groups $S_3$ and of various subgroups of it into the lepton sector of the standard model provides a good theoretical framework for explaining the
neutrinos masses and mixings as extracted from the experimental data. In this context we consider the see-saw mechanisms for two possible ansatze both in the presence of three Higgs doublets and of a Higgs triplet. This kind of scalar structure is natural in grand unification theories like SO(10) GUT.

In the first ansatz the $S_3$ group is implemented in the zeroth order in the neutrino sector, the $S_{12}$ subgroup in the first order and  the $S_{23}$ subgroup in the second order.
This case is realized in the presence of heavy right handed neutrinos with a mass around the GUT scale.

In the second ansatz again the $S_3$ group is implemented in the zeroth order, the $S_{23}$ subgroup in the first order and $S_{12}$ in the second.

The main difference occurs in the charged lepton sector where the left handed fields are arranged in the three dimensional representation for both cases whereas the left handed fields are triplets for the first ansatz and singlets for the second. Accordingly the three Higgs doublets are assigned to the $3 \times 3^*$ representation in the first ansatz and to the triplet in the second. The mixing matrix is the unit matrix in both cases with the correct masses for the first ansatz and with two massless charged leptons corresponding to the first two generations and one massive tau lepton for the second ansatz.

\section*{Acknowledgments} \vskip -.5cm
The work of R. J. was supported by PN 09370102/2009 and by a grant of the Ministry of National Education, CNCS-UEFISCDI, project number PN-II-ID-PCE-2012-4-0078. The work of J. S. was supported in part by the US DOE under Contract No. DE-FG-02-85ER 40231.

\end{document}